\documentclass[runningheads]{llncs}

\newif\ifdraft
\draftfalse

\usepackage{times}

\usepackage{todonotes}
\usepackage{xspace}
\usepackage{multirow}
\usepackage{hhline}
\usepackage{latexsym}
\usepackage{amssymb,amsmath}
\usepackage{url}
\usepackage{booktabs}
\usepackage{rotating}
\usepackage{paralist,enumerate}
\usepackage{verbatim}
\usepackage[inline]{enumitem}

\usepackage{graphicx}
\usepackage{caption}

\usepackage{stfloats}
\usepackage{float}

\newcommand{\shal}{\operatorname{ShallowWellbore}}

\newcommand{\well}{\operatorname{Wellbore}}

\newcommand{\name}{\operatorname{name}}
\newcommand{\comp}{\operatorname{completionYear}}

\ifdraft
\usepackage[ulem=normalem]{mychanges}
\else
\usepackage[final,ulem=normalem]{mychanges}
\fi
\setauthormarkup[right]{{\scriptsize[#1]}}
\definechangesauthor{DC}{blue}
\definechangesauthor{DL}{red}
\definechangesauthor{RW}{green}
\definechangesauthor{GX}{purple}

\definecolor{midgrey}{rgb}{0.5,0.5,0.5}
\definecolor{darkred}{rgb}{0.7,0.1,0.1}

\newcommand{\C}{\mathcal{C}}
\newcommand{\D}{\mathcal{D}}

\newcommand{\M}{\mathcal{M}}
\newcommand{\F}{\mathcal{F}}

\newcommand{\ignore}[1]{}

\newcommand{\tableRowBreak}{\\}
\newcommand{\tableColLargeSpace}{@{\hspace{2em}}}

\newcommand{\CalR}{\mathcal{R}}

\newcommand{\setB}[2]{\left\{ #1 \mathrel{}\middle|\mathrel{} #2\right\}}

\newcommand{\myPar}[1]{\medskip\noindent\textbf{#1}}
\newcommand{\col}{\operatorname{col}}
\newcommand{\ints}{\operatorname{ints}}

\begin{document}

\title{Data Scaling in OBDA Benchmarks. The VIG Approach.}

\author{Davide Lanti \and Guohui Xiao \and Diego Calvanese}

\institute{Free University of Bozen-Bolzano, Italy}
\maketitle

\begin{abstract}
In this paper we describe VIG, a data scaler for benchmarks in the
context of ontology-based data access (OBDA).  Data scaling is a relatively
recent approach, proposed in the database community, that allows for quickly
scaling up an input data instance to $s$ times its size, while preserving
certain application-specific characteristics. The advantage of the approach is
that the user is not required to manually input the characteristics of the
data to be produced, making it particularly suitable for OBDA benchmarks,
where the complexity of database schemas might pose a challenge for manual
input (e.g., the NPD benchmark contains 70 tables with some containing more
than 60 columns). As opposed to a traditional data scaler, VIG includes domain
information provided by the OBDA mappings and the ontology in order to produce
data.  VIG is currently used in the NPD benchmark, but it is not NPD-specific and can be seeded with
any data instance. The distinguishing features of VIG are
\begin{enumerate*}[label=\emph{(\arabic*)}]
\item its simple and clear generation strategy;
\item its efficiency, as each value is generated in constant time, without
  accesses to the disk or to RAM to retrieve previously generated values;
\item and its generality, as the data is exported in CSV files that can be
  easily imported by any RDBMS system.
\end{enumerate*}

 VIG is a java implementation licensed under Apache~2.0, and its source code is
 available on GitHub (https://github.com/ontop/vig) in the form of a Maven project. The code is being maintained since two years by the -ontop- team at the Free University of
 Bozen-Bolzano.

\end{abstract}

\section{Introduction}

An important research problem in Big Data is how to provide end-users with
transparent access to the data, abstracting from storage details.  The paradigm
of Ontology-based Data Access (OBDA)~\cite{CDLL*06} provides an answer to this
problem that is very close to the spirit of the Semantic Web. In OBDA the data
stored in a relational database is presented to the end-users as a
\emph{virtual} RDF graph over which SPARQL queries can be posed. This solution
is realized through \emph{mappings} that link classes and properties in the
ontology to queries over the database.

Proper benchmarking of query answering systems, such as OBDA systems, requires
scalability analyses taking into account data instances of increasing
volume. Such instances are often provided by generators of synthetic
data. However, such generators are either complex ad-hoc implementations
working for a specific schema, or require considerable manual input by the
end-user. The latter problem is exacerbated in the OBDA setting, where database
schemas tend to be particularly big and complex (e.g., 70 tables, some with
more of 80 columns in~\cite{LRXC15}). The result of having
\replaced[DC]{only}{a} few benchmarks is that \replaced[DC]{they}{benchmarks}
become increasingly misused over time. For instance, evaluations on OBDA
systems are usually performed on benchmarks originally designed to test triple
stores, although the two systems are totally different and present different
bottlenecks~\cite{LRXC15}.

Data scaling~\cite{Tay20131168} is a recent approach that tries to overcome
this problem by automatically tuning the generation parameters through
statistics collected over an initial data instance. Hence, the same generator
can be reused in different contexts, as long as an initial data instance is
available. A measure of quality for the produced data is defined in terms of
results for the \replaced[DC]{available queries}{queries available}, that
should be \emph{similar} to the one observed for real data of comparable
volume. In the context of OBDA, taking as\added[DC]{ the} only parameter for
generation an initial data instance does not produce data of acceptable
quality, since it has to comply with constraints deriving from the structure of
the mappings and the ontology, that in turn derive from the
\replaced[DC]{application domain}{domain of application}.

In this work we present VIG, a data scaler for OBDA benchmarks. In the
VIG system, we lift the scaling approach from the instance level to
the OBDA level, where the domain information of ontologies and mappings
has to be taken into account as well. VIG is extremely efficient and
suitable to generate huge amounts of data, as tuples are generated in
constant time without disk accesses or need to retrieve previously
generated values.  Furthermore, different instances of VIG can be
delegated to different machines, and parallelization can scale up to
the number of columns in the schema, without communication overhead.


VIG is a Java implementation licensed under Apache 2.0, and its source code is
available on GitHub in the form of a Maven project~\cite{vig}. The code is
maintained by the Ontop team at the \added[DC]{Free }University of
Bozen-Bolzano, and it comes with extensive documentation in\added[DC]{ the}
form of Wiki pages.

The rest of the paper is structured as follows. In
Section~\ref{s:basics}, we introduce the basic notions and notation to
understand this paper. In Section~\ref{s:measures}, we define the
scaling problem and discuss important measures on the produced data
that define the quality of instances in a given OBDA setting.  In
Section~\ref{s:algorithm}, we discuss the VIG algorithm, and how it
ensures that data conforming to the identified measures is
produced. Sections~\ref{s:related} and~\ref{s:conclusion} contain
related work and conclusions, respectively.

\section{Basic Notions and Notation}
\label{s:basics}

We \replaced[DC]{assume}{take for granted} that the reader has moderate
knowledge of OBDA\replaced[DC]{, and refer for it}{. If this is not the case,
 then the reader should refer} to the abundant literature on the subject,
like~\cite{CDLL*09}. Moreover, \replaced[DC]{we assume familiarity}{the reader
 should be familiar} with basic notions from probability calculus and
statistics.

The W3C standard ontology language in OBDA is
OWL~2~QL~\cite{W3Crec-OWL2-Profiles}. For the sake of conciseness, we
\replaced[DC]{consider here}{here consider} its mathematical underpinning
\textit{DL-Lite}$_\CalR$~\cite{CDLLR07}. Table~\ref{t:onto} shows a portion of
the ontology from the NPD benchmark, which is the foundation block of our
running example.

The W3C standard query language in OBDA is
SPARQL~\cite{W3Crec-SPARQL-1.1-query}, with queries evaluated under the
OWL~2~QL entailment regime~\cite{KRRXZ14}. Intuitively, under this semantics
each basic graph pattern (BGP) can be seen as a single conjunctive
query\added[DC]{ (CQ)} \emph{\replaced[DC]{without}{devoid of} existentially
 quantified variables}. As in our examples we will only refer to SPARQL queries
containing exactly one BGP, we will use the more concise syntax for
\replaced[DC]{CQs}{conjunctive queries} rather than the SPARQL
syntax. Table~\ref{t:queries} contains the queries that we will consider in our
running example.

The mapping component links predicates in the ontology to queries over the
underlying relational database. To present our techniques, we need to introduce
this component in a formal way. The standard W3C syntax for mappings is
R2RML~\cite{W3Crec-R2RML}, however here we use a more concise syntax that is
common in the OBDA literature. Formally, a \emph{mapping assertion}
$\mathfrak{m}$ is an expression of the form
$X(\vec{f}, \vec{x}) \leftarrow \operatorname{conj}(\vec{y})$,
consisting\added[DC]{ of} a \emph{target} part
$X(\vec{f}, \vec{x})$\replaced[DC]{, which}{ that} is an atom over function
symbols $\vec{f}$ (also called \emph{templates}) and variables
$\vec{x} \subseteq \vec{y}$, and a \emph{source} part
$\operatorname{conj}(\vec{y})$\replaced[DC]{, which}{ that} is a
\replaced[DC]{CQ}{conjunctive query} whose output variables are $\vec{y}$. We
say that $\mathfrak{m}$ \emph{defines the predicate $X$} if $X$ is in the
target of $\mathfrak{m}$. A \emph{basic mapping} is a mapping whose source part
contains exactly one atom. Table~\ref{t:map} contains the mappings for our
running example, as well as a short description of how these mappings are used
in order to create a (virtual) set of \emph{assertions}.

For the rest of this paper we fix an \emph{OBDA instance}
$(\mathcal{O}, \mathcal{M}, \Sigma, \D)$, where $\mathcal{O}$ is an OWL~2~QL
ontology, $\Sigma$ is a database schema with foreign and primary key
dependencies, $\mathcal{M}$ is a set of mappings linking predicates in
$\mathcal{O}$ to queries over $\Sigma$, and $\D$ is a database instance that
satisfies the dependencies in $\Sigma$ and the disjointness axioms in
$\mathcal{O}$.  We denote by $\col(\Sigma)$ the set of all columns in $\Sigma$.
Given a column $C \in \col(\Sigma)$, we denote by $C^\D$ the set of values for
$C$ in $\D$. Finally, given a term $f(\vec{x})$,\added[DC]{ where}
$\vec{x} = (x_1,\ldots,x_{p},\ldots,x_n)$, we denote the argument $x_p$
at position $p$ by $f(\vec{x})|_p$.

\begin{table}[t]
\centering
\scriptsize
\caption{Portion of the ontology for the NPD benchmark. The
 first three axioms (left to right) state that the classes ``DevelopmentWellbore'',
 ``ExplorationWellbore'', and ``SuspendedWellbore'' are subclasses of the class
 ``Wellbore''. The fourth axiom states that the classes ``ExplorationWellbore''
 and ``DevelopmentWellbore'' are disjoint.}
\label{t:onto}
\begin{tabular}{l \tableColLargeSpace l}
  \toprule
  DevelopmentWellbore $\sqsubseteq$ Wellbore & ExplorationWellbore $\sqsubseteq$ Wellbore \tableRowBreak
  SuspendedWellbore $\sqsubseteq$ Wellbore & ExplorationWellbore $\sqcap$ DevelopmentWellbore $\sqsubseteq \bot$ \tableRowBreak
  \bottomrule
\end{tabular}
\end{table}

\begin{table}[t]
  \centering
  \scriptsize
  \caption{\label{t:queries} Queries for our running example.}
  \begin{tabular}{l l}
    \toprule
    $q_1(y)$ & $\leftarrow \well(y), \operatorname{developmentWellboreForField}(x,y)$ \tableRowBreak
    $q_2(x,n,y)$ & $\leftarrow \well(x), \name(x,n), \comp(x,y)$ \tableRowBreak
    \bottomrule
  \end{tabular}
\end{table}

\begin{table}[t]
  \centering
  \scriptsize
  \caption{\label{t:map}Mappings from the NPD benchmark. Results from the evaluation of the queries on the source part build predicates in the ontology. For example, each triple $(a,b,c)$ in a relation for \texttt{development\_wellbores} corresponds to a predicate $\shal(w(a))$ in the ontology. In the R2RML mappings for the original NPD benchmark the term $w(id)$ corresponds to the URI template \texttt{npd:wellbore/\{id\}}. Columns named \texttt{id} are primary keys, and the column \texttt{fid} in \texttt{development\_wellbores} is a foreign key for the primary key \texttt{fid} of the table \texttt{fields}.}
  \begin{tabular}{l l}
    \toprule
    $\operatorname{DevelopmentWellbore}(w(id))$ & $\leftarrow$ \texttt{development\_wellbores(id,name,year,fid)}\tableRowBreak
    $\operatorname{ExplorationWellbore}(w(id))$ & $\leftarrow$ \texttt{exploration\_wellbores(id,name,year,state)}\tableRowBreak
    $\operatorname{SuspendedWellbore}(w(id))$ & $\leftarrow$ \texttt{exploration\_wellbores(id,name,year,state),}\tableRowBreak
    &~~~~~~\texttt{state='suspended'}\tableRowBreak
    $\operatorname{Field}(f(fid))$ & $\leftarrow$ \texttt{fields(fid,name)} \tableRowBreak
   $\operatorname{completionYear}(w(id),year)$ & $\leftarrow$ \texttt{\texttt{development\_wellbores(id,name,year,fid)}} \tableRowBreak
    $\operatorname{name}(w(id),name)$ & $\leftarrow$ \texttt{\texttt{development\_wellbores(id,name,year,fid)}} \tableRowBreak
    $\operatorname{completionYear}(w(id),year)$ & $\leftarrow$ \texttt{\texttt{exploration\_wellbores(id,name,year)}} \tableRowBreak
    $\operatorname{name}(w(id),name)$ & $\leftarrow$ \texttt{exploration\_wellbores(id,name,year)} \tableRowBreak
    $\operatorname{developmentWellboreForField}(w(id), f(fid)) $ & $\leftarrow$ \texttt{development\_wellbores(id,name,year,fid),} \tableRowBreak
    &~~~~~~\texttt{fields(fid,fname)} \tableRowBreak

    \bottomrule
  \end{tabular}
\end{table}

\section{Data Scaling for OBDA Benchmarks: VIG Approach}
\label{s:measures}

The \emph{data scaling problem} introduced in~\cite{Tay20131168} is formulated as follows:

\begin{definition}[\textbf{Data Scaling Problem}]
  Given a dataset $\D$, produce a dataset $\D'$ which is \emph{similar} to $\D$
  but \replaced[DC]{$s$}{$n$} times its size.
\end{definition}

The notion of \emph{similarity} is application-based. Being our goal
benchmarking, we define similarity in terms of query results for the queries at
hand. In~\cite{Tay20131168}, \added[DC]{the }authors do not consider such
queries to be available to the generator, since their goal is broader than
benchmarking over a pre-defined set of queries. In OBDA benchmarking, instead,
the (SQL) workload for the database can be estimated from the mapping
component. Therefore, VIG includes the mappings in the analysis, so as to
obtain a more realistic and OBDA-tuned generation.

Concerning the size, similarly to other approaches, VIG scales each table in
$\D$ by a factor of $s$.

\subsection{Similarity Measures for OBDA and Their Rationale}

We overview the similarity measures used by VIG, and why they are important in
the scenario of OBDA benchmarking.


\myPar{Schema Dependencies.}  $\D'$ should be a valid instance for
$\Sigma$. VIG is, to the best of our knowledge, the only data scaler
able to generate in constant time tuples that satisfy multi-attribute
primary keys for \emph{weakly-identified entities}\footnote{In a
  relational database, a weak entity is an entity that cannot be
  uniquely identified by its attributes alone.}. The current
implementation of VIG does not support multi-attribute foreign keys.

\myPar{Column-based Duplicates and NULL Ratios.}  They respectively measure the
ratio of duplicates and \added[DC]{of }nulls in a given column, and are common
parameters for the cost estimation performed by query planners in databases. By
default, VIG maintains them in $\D'$ to preserve the cost of joining columns in
a key-foreign key relationship (e.g., the join from the last mapping in our
running example). This default behavior, however, is not applied with
\emph{fixed-domain} columns, \replaced[DC]{which}{that} are columns whose
content does not depend on the size of the database instance. The column
\texttt{state} in the table \texttt{exploration\_wellbore} is fixed-domain,
because it partitions the elements of \texttt{id} into a fixed number of
classes\footnote{The number of classes in the ontology does not depend on the
 size of the data instance.}. VIG analyzes the mappings to detect fixed-domain
columns, and additional fixed-domain columns can be manually specified by the
user. To generate values for a fixed-domain column, VIG reuses the values found
in $\D$ so as to prevent empty answers for the SQL queries in the mappings. For
instance, a value `\texttt{suspended}' must be generated for the column
\texttt{state} in order to produce objects for the class SuspendedWellbore.

VIG generates values in columns according to a \emph{uniform distribution},
that is, values in columns have all the same probability of being
repeated. Replication of the distributions from $\D$ will be included the next
releases of VIG.

\myPar{Size of Columns Clusters, and Disjointness.}  Query $q_1$ from our
running example returns an empty set of answers, regardless\added[DC]{ of} the
considered data instance. This\added[DC]{ is} because the function $w$ used to
build objects for the class Wellbore does not match with the function $f$ used
to build objects for Fields. Indeed, fields and wellbores are two different
entities for which a join operation would be meaningless.

On the other hand, a standard OBDA translation of $q_2$ into SQL produces a
\replaced[DC]{union of CQs}{UCQ} containing several joins between the two tables
\texttt{development\_wellbores} and \texttt{exploration\_wellbores}. This is
possible only because the mappings for Wellbore, name, and completionYear all
use the \emph{same} unary function symbol $w$ to define wellbores. Intuitively,
every pair of terms over the same function symbol and appearing on the target
of two distinct basic mappings identifies sets of columns for which the join
operation is semantically meaningful\footnote{Therefore, for which a join could
 occur during the evaluation of a user query.}. Generating data that guarantees
the correct cost for these joins is crucial in order to deliver a realistic
evaluation. In our example, the join between \texttt{development\_wellbore} and
\texttt{exploration\_wellbore} over \texttt{id} is empty under $\D$ (because
ExplorationWellbore and DevelopmentWellbore are disjoint classes). VIG is able to
replicate this fact in $\D'$. This implies that VIG can generate data satisfying
disjointness constraints declared over classes whose individuals are
constructed from a unary template in a basic mapping, if $\D$ satisfies those
constraints.

\section{The VIG Algorithm}
\label{s:algorithm}

We now show how VIG realizes the measures described in the previous section.
The building block of VIG is a \emph{pseudo-random number generator}, that is a
sequence of integers $(s_i)_{i \in \mathbb{N}}$ defined through a transition
function $s_k := f(s_{k-1})$. \replaced[DC]{The authors}{Authors}
in~\cite{export:68246} discuss a particular class of pseudo-random generators
based on \emph{multiplicative groups modulo a prime number}. Let $n$ be the
number of distinct values to generate. Let $g$ be a generator for the
multiplicative group modulo a prime number $p$, with $p > n$. Consider the
sequence
$S:= \langle g^i~mod~p\ |\ i = 1, \ldots, p \text{ and } (g^i~mod~p) \le n
\rangle$. Then $S$ is a \emph{permutation} of values in the interval
$[1, \ldots, n]$. \replaced[DC]{Here}{In this section} we show how this
generator is used in VIG to quickly produce data \replaced[DC]{complying}{that
 complies} with foreign and primary key constraints.

From now on, let $s$ be a scale factor, and let
$\operatorname{dist}(C,\D)$ denote the number of distinct non-null
values in a column $C$ in the database instance $\D$. Let
$\operatorname{size}(T,\D)$ denote the number of tuples occurring in
the table $T$ in the database instance $\D$.
For each column~$c$, VIG creates a set of intervals $\ints(c)$ and generates
values accordingly.


\myPar{Initialization Phase.} For each table $T$, VIG sets the number
$\operatorname{size}(T,\D')$ of tuples to generate to
$\operatorname{size}(T,\D) * s$. Then, VIG calculates the number of non-null
distinct values that need to be generated for each column, given $s$ and
$\D$. That is, for each column $C$, if $C$ is not fixed-domain then VIG sets
$\operatorname{dist}(C,\D') := \operatorname{dist}(C,\D) * s$. Otherwise, $\operatorname{dist}(C,\D')$ is set to
$\operatorname{dist}(C,\D)$.

\myPar{Creation of Intervals.}  When $C$ \replaced[DC]{is}{be} a numerical
column, VIG initializes $\ints(C)$ by the interval
$I_C := [min(C,\D), min(C,\D) + \operatorname{dist}(C,\D')-1]$ of distinct values
to be generated, where $min(C,\D)$ denotes the minimum value occurring in
$C^\D$. Otherwise, if $C$ is non-numerical, $\ints(C)$ is initialized to the
interval $I_C := [1, \operatorname{dist}(C,\D')]$.  The elements in $\ints(C)$
will be transformed into values of the desired datatype by \replaced[DC]{a
 suitable}{some} injective function in the final \replaced[DC]{generation
 step}{step of generation}.

\myPar{Primary Keys Satisfaction.}
Let $K = \{C_1, \ldots, C_n\}$ be the primary key of a table $T$. 
In order to ensure that values generated for each column through the pseudo-random
generator will not lead to duplicate tuples in $K$, the least common multiple
$\operatorname{lcm}(\operatorname{dist}(C_1,\D'), \ldots,
\operatorname{dist}(C_n,\D'))$ must be greater than
$\operatorname{tuples}(T,\D')$. If this is not true, then VIG ensures the condition by slightly increasing $\operatorname{dist}(C_i,\D')$ for some column $C_i$ in
$K$. Once the condition holds, data can be generated independently for each column without risk of generating duplicate tuples for $K$.

\myPar{Columns Cluster Analysis.}
In this phase, VIG analyzes $\M$ in order to identify columns that
could be joined in a translation to SQL, and groups them together in
\emph{pre-clusters}. Formally, let
$X_1(\vec{f_1}, \vec{x_1}), \ldots, X_m(\vec{f_m}, \vec{x_m})$ be the
atoms defined by basic mappings in $\M$. Let
$\F = \bigcup_{i = 1 \ldots m} \setB{f(\vec{x})}{f(\vec{x})\text{ is a term in }
  X_i(\vec{f_i}, \vec{x_i})}$ be the set of all the terms occurring in
such atoms. A set of columns $\mathfrak{pc}$ is a \emph{pre-cluster}
if there exists a function $f$ and a valid position $p$ in $f$ such
that
$\mathfrak{pc} = \setB{f(\vec{x})|_p}{f(\vec{x}) \in
  \mathcal{F}}$.

VIG evaluates on $\D$ all combinations of such joins between columns in
a pre-cluster $\mathfrak{pc}$, and produces values in $\D'$ so that the
selectivities for these joins are maintained. In order to do so, the
intervals for the columns in $\mathfrak{pc}$ are modified. This
modification must be propagated to all the columns related via a
foreign key relationship to some column in $\mathfrak{pc}$. In
particular, the modification might propagate up to columns belonging
to different pre-clusters, inducing a clash. VIG groups together such
pre-clusters in order to avoid this issue. Formally, let
$\mathcal{PC}$ denote the set of pre-clusters for $\M$. Two
pre-clusters $\mathfrak{pc}_1, \mathfrak{pc}_2 \in \mathcal{PC}$ are
in \emph{merge relation}, denoted as
$ \mathfrak{pc}_1 \leftrightsquigarrow \mathfrak{pc}_2$, iff
$\mathcal{C}(\mathfrak{pc}_1) \cap \mathcal{C}(\mathfrak{pc}_2) \neq
\emptyset$, where
$\mathcal{C}(\mathfrak{pc}) = \{D\in\col(\Sigma)\mid \text{there is a
  } C \in \mathfrak{pc} : D \overset{*}{\leftrightarrow} C\}$, where
$\overset{*}{\leftrightarrow}$ is the reflexive, symmetric, and
transitive closure of the single column foreign key relation between
pairs of columns\footnote{Remember that VIG does not allow for
  multi-attribute foreign keys.}.  Given a pre-cluster $\mathfrak{pc}$,
the set of columns
$\{c \in \mathfrak{pc'} \mid \mathfrak{pc'}
{\overset{*}{\leftrightsquigarrow}} \mathfrak{pc} \}$
is called a \emph{columns cluster}, where
$\overset{*}{\leftrightsquigarrow}$ is the transitive closure of
$\leftrightsquigarrow$.
Columns clusters group together those pre-clusters for which columns cannot be generated independently.


After identifying columns clusters, VIG analyzes the number of shared elements
between the columns in the cluster, and creates new intervals
accordingly. Formally, consider the columns cluster
$\mathfrak{cc}$. Let
$H \subseteq \mathfrak{cc}$ be a set of columns, and the set 
$\mathcal{K}_H:=\setB{C}{C \in K, H \subset K \subseteq \mathfrak{cc}}$ 
of columns in the super-sets of $H$. For each such $H$, VIG creates an interval
$I_H$ such that
$|I_H| := |\bigcap_{C \in H} C^\D \setminus \bigcap_{C \in \mathcal{K}_H} C^\D| *
s$,
and adds $I_H$ to $\ints(C)$ for all $C\in H$.  Boundaries for
all intervals $I_H$ are set in a way that they do not overlap.


\myPar{Foreign Keys Satisfaction.} At this point, foreign key columns $D$ for
which there is no columns cluster $\mathfrak{pc}$ such\added[DC]{ that}
$D \in \mathcal{C}(\mathfrak{pc})$, have a single interval whose boundaries
have to be aligned to the (single) interval of the parent. Foreign keys
relating pairs of columns in a cluster, instead, are already satisfied by
construction of the intervals in the columns cluster. More work, instead, is
necessary for columns belonging to
$\mathcal{C}(\mathfrak{cc}) \setminus \mathfrak{cc}$, for some columns cluster
$\mathfrak{cc}$. VIG encodes the problem of finding intervals for these columns
that satisfy the number of distinct values and the foreign key constraints into
a \emph{constraint program} (see Table~\ref{t:choco}), which is solved by an
off-the-shelf constraint solver, e.g., Choco~\cite{choco}.

\begin{table}
  \centering
  \scriptsize
  \caption{\label{t:choco} CSP Program for the Choco Solver. In the following, $S$ is the set of intervals for the columns in the columns cluster $\mathfrak{cc}$, plus one extra disjoint interval. Each interval $I$ in a column $C$ is encoded as a pair of variables $X_{\langle C,I\rangle},Y_{\langle C,I \rangle}$, keeping respectively the lower and upper limit for the interval.}
  \begin{tabular}{l}
    \toprule
    Create Program Variables: \tableRowBreak
    $\forall I \in S.\ \forall C \in \C(\mathfrak{cc}).\ X_{\langle C,I\rangle},Y_{\langle C,I \rangle} \in [I.min, I.max]$ \tableRowBreak
    Set Boundaries for Known Intervals: \tableRowBreak
    $\forall I \in S.\ \forall C \in \C(\mathfrak{cc}).\ I \in \ints(C) \Rightarrow X_{\langle C,I \rangle} = I.min, Y_{\langle C,I \rangle} = I.max$ \tableRowBreak
    Set Boundaries for Known Empty Intervals: \tableRowBreak
    $\forall I \in S.\ \forall C \in \mathfrak{cc}.\ I \notin \ints(C) \Rightarrow X_{\langle C,I \rangle} = Y_{\langle C,I \rangle}$ \tableRowBreak
    The Y's should be greater than the X's: \tableRowBreak
    $\forall I \in S.\ \forall C \in \C(\mathfrak{cc}).\ X_{\langle C, I \rangle} \le Y_{\langle C, I \rangle}$ \tableRowBreak
    Foreign Keys (denoted by $\subseteq$): \tableRowBreak
    $\forall I \in S.\ \forall C_1 \in (\C(\mathfrak{cc}) \setminus \mathfrak{cc}).\ \forall C_1 \subseteq C_2.\ X_{\langle C_1, I \rangle} \ge X_{\langle C_2, I \rangle}$ \tableRowBreak
    $\forall I \in S.\ \forall C_1 \in (\C(\mathfrak{cc}) \setminus \mathfrak{cc}).\ \forall C_2 \subseteq C_1.\ X_{\langle C_2, I \rangle} \ge X_{\langle C_1, I \rangle}$ \tableRowBreak
    $\forall I \in S.\ \forall C_1 \in (\C(\mathfrak{cc}) \setminus \mathfrak{cc}).\ \forall C_1 \subseteq C_2.\ Y_{\langle C_1, I \rangle} \le Y_{\langle C_2, I \rangle}$ \tableRowBreak
    $\forall I \in S.\ \forall C_1 \in (\C(\mathfrak{cc}) \setminus \mathfrak{cc}).\ \forall C_2 \subseteq C_1.\ Y_{\langle C_2, I \rangle} \le Y_{\langle C_1, I \rangle}$ \tableRowBreak
    Width of the Intervals: \tableRowBreak
    $\sum_{C,I} Y_{\langle C, I \rangle} - X_{\langle C, I \rangle} = |C|$ \tableRowBreak
    \bottomrule
  \end{tabular}
\end{table}
\myPar{Generation.} At this point, each column in $\col(\Sigma)$ is associated
to a set of intervals. The elements in the intervals are associated to values
in the column datatype, and to values from $C^\D$ in case $C$ is
fixed-domain. VIG uses the pseudo-random number generator to randomly pick
elements from the intervals that are then transformed into database
values. NULL values are generated according to the detected NULLS
ratio. Observe that the generation of a value in a column takes constant time
and can happen independently for each column, thanks to the previous phases in
which intervals were calculated.

\section{Related Work}
\label{s:related}

UpSizeR~\cite{Tay20131168} replicates two kinds of distributions observed on
the values for the key columns, called \emph{joint degree distribution} and
\emph{joint distribution over co-clusters}\footnote{The notion of co-cluster
 has nothing to do with the notion of columns-cluster introduced
 here.}. However, this requires several assumptions to be made on the $\Sigma$,
for instance tables can have at most\deleted[DC]{ than} two foreign keys,
primary keys cannot be multi-attribute, etc. Moreover, generating values for
the foreign keys require reading of previously generated values, which is not
required in VIG. A strictly related approach is \emph{Rex}~\cite{rex-2015},
which provides, through the use of dictionaries, a better handling of the
content for non-key columns.

In terms of similarity measures, the approach \replaced[DC]{closest}{closer} to
VIG is \emph{RSGen}~\cite{RevStats}, that also
\replaced[DC]{considers}{consider} measures like \texttt{NULL} ratios or number
of distinct values. Moreover, values are generated according to a uniform
distribution, as in VIG. However, the approach only works on numerical data
types, and it seems not to \replaced[DC]{support}{provide support for}
multi-attribute primary keys.

In \emph{RDF graph scaling}~\cite{Qiao:2015:RAR:2723372.2746479}, an
additional parameter, called \emph{node degree scaling factor}, is
provided as input to the scaler. The approach is able to replicate the
phenomena of \emph{densification} that have been observed for certain
types of networks. We see this as a meaningful extension for VIG, and
we are currently studying the problem of how this could be applied in
an OBDA context.

Observe that all the approaches above do not consider ontologies nor
mappings. Therefore, many measures important in a context with
mappings and ontologies and discussed here, like selectivities for
joins in a co-cluster, class disjointness, or reuse of values for
fixed-domain columns, cannot be handled by any of them.

\section{Conclusion and Development Plan}
\label{s:conclusion}

In this work we presented VIG, a data-scaler for OBDA benchmarks. VIG
integrates some of the measures used by database query optimizers and existing
data scalers with OBDA-specific measures, in order to deliver a better data
generation in the context of OBDA benchmarks. VIG is available as a Java maven
project on GitHub, and it comes with extensive documentation in form of wiki
pages. VIG is a mature implementation that is being delivered since two years
together with the NPD benchmark. VIG is licensed under Apache~2.0, and is
maintained at the Free University of Bozen-Bolzano. It is extremely efficient
and suitable to generate huge amounts of data. In our experience, VIG can
generate hundreds of Gigabytes in \replaced[DC]{just}{a matter of} a few hours
on a normal laptop.
The current work plan is to enrich the quality of the data produced by adding
support for multi-attribute foreign keys, joint degree and value distributions,
and intra-row correlations (e.g., objects from SuspendedWellbore might not have
a completionYear). Unfortunately, it can be proved that some of these measures
conflict with the current feature of constant time for generation of
tuples. Moreover, many of them require access to previously generated tuples in
order to be calculated (e.g., joint-degree distribution~\cite{Tay20131168}).

\bibliographystyle{splncs03}
\bibliography{main}

\end{document}